\documentclass[10pt,prl,aps,twocolumn,nofootinbib]{revtex4-1}
\usepackage{amssymb,amsmath,amsfonts,amsbsy,epsfig,graphicx}
\usepackage[dvipsnames]{xcolor}
\usepackage{braket}
\definecolor{lgray}{gray}{0.35}
\definecolor{cor}{rgb}{0.94, 0.41, 0.35}
\definecolor{cas}{rgb}{0.63, 0.72, 0.83}
\usepackage[colorlinks=true,urlcolor=Gray,linkcolor=lgray,
citecolor=Gray,
             pdfpagelabels=true,hypertexnames=true,
            plainpages=false,naturalnames=false,
             ]{hyperref}
\usepackage[utf8]{inputenc}

\newcommand{\be}{\begin{equation}}
\newcommand{\ee}{\end{equation}}
\newcommand{\bea}{\begin{eqnarray}}
\newcommand{\eea}{\end{eqnarray}}
\newcommand{\nn}{\nonumber}

\newcommand{\de}{{\rm d}}
\newcommand{\K}{{\boldsymbol k}}
\newcommand{\x}{{\boldsymbol x}}
\newcommand{\q}{{\boldsymbol q}}
\DeclareMathAlphabet\mathbfcal{OMS}{cmsy}{b}{n}
\begin{document}

\title{Seeding primordial black holes in multif{i}eld inf{l}ation}

\date{July 25, 2020}

\author{
Gonzalo A. Palma$^{a}$, Spyros Sypsas$^{b}$ and Cristobal Zenteno$^{a}$
}

\affiliation{
${}^{a}$Grupo de Cosmolog\'ia y Astrof\'isica Te\'orica, Departamento de F\'{i}sica, FCFM, \mbox{Universidad de Chile}, Blanco Encalada 2008, Santiago, Chile\\
${}^{b}$Department of Physics, Faculty of Science, Chulalongkorn University, Phayathai Rd., Bangkok 10330, Thailand}

\begin{abstract}

The inflationary origin of primordial black holes (PBHs) relies on a large enhancement of the power spectrum $\Delta_\zeta$ of the curvature fluctuation $\zeta$ at wavelengths much shorter than those of the cosmic microwave background anisotropies. This is typically achieved in models where $\zeta$ evolves without interacting significantly with additional (isocurvature) scalar degrees of freedom. However, quantum gravity inspired models are characterized by moduli spaces with highly curved geometries and a large number of scalar fields that could vigorously interact with $\zeta$ (as in the cosmological collider picture). Here we show that isocurvature fluctuations can mix with $\zeta$ inducing large enhancements of its amplitude. This occurs whenever the inflationary trajectory experiences rapid turns in the field space of the model leading to amplifications that are exponentially sensitive to the total angle swept by the turn, which induce characteristic observable signatures on $\Delta_\zeta$. We derive accurate analytical predictions and show that the large enhancements required for PBHs demand noncanonical kinetic terms in the action of the multifield system.

\end{abstract}

\maketitle


{\bf Introduction.} Unlike their astrophysical counterparts, primordial black holes (PBHs)~\cite{Hawking:1971ei,Carr:1974nx} might have stemmed from large statistical excursions of the primordial curvature fluctuation $\zeta$ generated during the pre-Big-Bang era known as cosmic inflation~\cite{Guth:1980zm,Starobinsky:1980te,Linde:1981mu,Albrecht:1982wi,Mukhanov:1981xt}. Once inflation is over, these fluctuations can induce overdense regions with matter that, collapsing under the pull of gravity, give birth to PBHs. In the correct abundance, these provide an excellent dark matter candidate~\cite{Ivanov:1994pa, Carr:2016drx, Inomata:2017okj, Georg:2017mqk, Kovetz:2017rvv}, braiding early and late Universe dynamics.

Cosmic microwave background (CMB) observations over the range of scales $10^{-4}\phantom{l}{\rm Mpc}^{-1} \lesssim k \lesssim 10^{-1}\phantom{l}{\rm Mpc}^{-1}$ confirm that, after inflation, $\zeta$ was distributed according to a Gaussian statistics determined by a nearly scale invariant power spectrum $\Delta_\zeta (k)$~\cite{Akrami:2018odb, Akrami:2019izv}. If these properties persisted all the way up to the smallest cosmological wavelengths, large fluctuations of $\zeta$ would constitute extremely rare events, preventing PBH formation. Inflationary PBHs, with an abundance compatible with dark matter~\cite{Carr:2020gox}, thus require a strong scale dependence at $ k \gtrsim 10^{8}\phantom{l}{\rm Mpc}^{-1}$, through an amplification of $\Delta_\zeta (k)$ at least $10^{7}$ larger than its CMB value. 

Such an amplification can be achieved in several ways. Examples include: single-field models with special potentials~\cite{Leach:2000ea,Alabidi:2009bk,Kohri:2007qn,Garcia-Bellido:2017mdw,Germani:2017bcs,Motohashi:2017kbs,Hertzberg:2017dkh}; single-field models with resonant backgrounds~\cite{Cai:2018tuh,Cai:2019bmk}; models with light spectator fields~\cite{Yokoyama:1995ex, Kawasaki:2012wr,Kohri:2012yw,Pi:2017gih}; models where the inflaton couples to gauge fields~\cite{Linde:2012bt, Bugaev:2013fya}; models with resonant instabilities during the pre-heating inflaton's decay~\cite{Green:2000he,Bassett:2000ha,Martin:2019nuw}. In this work, we are particularly interested in re-examining the enhancement of $\Delta_\zeta (k)$ within the paradigm of multifield inflation~\cite{Sakharov:1993qh,Randall:1995dj, GarciaBellido:1996qt, Kawasaki:1997ju, Clesse:2015wea}. Ultraviolet complete frameworks (such as supergravity and string theory) lead to models with a variety of fields charting multi-dimensional target spaces with curved geometries. The effective field theory description of these models, valid during inflation, includes potentially sizable interactions between $\zeta$ and other (isocurvature) fluctuations~\cite{Chen:2009zp, Cremonini:2010ua, Achucarro:2010da,Pi:2012gf}, an idea that, recently, has drawn considerable attention through the cosmological collider program~\cite{Noumi:2012vr,Arkani-Hamed:2015bza, Chen:2015lza}. 

The purpose of this letter is to show that a purely multifield mechanism can indeed lead to enhancements of $\Delta_{\zeta} (k)$ large enough to produce PBHs abundantly. In multifield inflation, the leading interaction between $\zeta$ and isocurvature fluctuations is proportional to the angular velocity $\Omega$ with which the inflationary trajectory experiences turns in the target space. The total angle swept by the turning trajectory is of order $\delta \theta \sim \Omega \delta t$, where $\delta t$ is the duration of turn. Remarkably, we derive an accurate analytical prediction for $\Delta_{\zeta} (k)$ generated by sudden turns, valid in the rapid turn regime~\cite{Cremonini:2010ua, Achucarro:2010da, Cespedes:2012hu, Achucarro:2012yr, Assassi:2013gxa, Brown:2017osf, Iyer:2017qzw, An:2017hlx, Achucarro:2019pux, Fumagalli:2019noh, Bjorkmo:2019fls, Christodoulidis:2019jsx}, where $\Omega$ is much greater than $H$, the Hubble expansion rate during inflation. This allows us to show that under a turn of total angle $\delta \theta$ of order $1$ or larger, the power spectrum is enhanced exponentially as
\be
\Delta_\zeta \sim  \frac{ e^{2 \delta \theta} }{4(1 + 4 \delta \theta^2)} \times \Delta_{\zeta , {\rm CMB}} , \label{main-result}
\ee
where $ \Delta_{\zeta , {\rm CMB}}$ is its amplitude at CMB scales. Therefore, in order to have a large enhancement of around $10^{7}$, the total swept angle $\delta \theta$ must be about $4 \pi$. Since models with canonical kinetic terms must respect  $\delta \theta < \pi$, we conclude that large enhancements can only be achieved in models with curved target spaces (noncanonical kinetic terms) as often encountered in ultraviolet complete theories. Thus, our proposal opens up a new window of opportunity to test the existence of additional degrees of freedom interacting with $\zeta$ during inflation.


{\bf Multifield inflation.} To set the context, we review some general aspects of multifield inflation in the particular case of two fields. We consider a general action of the form
\be
S =  S_{\rm EH}  - \!\! \int \! \de^4 x \sqrt{-g} \left[  \frac{1}{2} \gamma_{ab} (\phi)  \, \partial_{\mu} \phi^a \partial^{\mu} \phi^b + V (\phi) \right] , \label{action-starting_point}
\ee
where $S_{\rm EH}$ is the Einstein-Hilbert term constructed from a spacetime metric $g_{\mu \nu}$ with determinant $g$ and $\gamma_{a b}$ is a metric characterizing the geometry of the target space spanned by the fields $\phi^a = (\phi^1 , \phi^2)$. 

Spatially flat cosmological backgrounds are described by the line element $\de s^2 = - \de t^2 + a^2 \de {\x}^2$, where $a = a(t)$ is the usual scale factor. In these spacetimes, the background scalar fields $\phi_0^a(t)$ satisfy:
\be
D_t \dot \phi_0^a + 3 H \dot \phi_0^a + \gamma^{ab} V_b (\phi_0) = 0 , \label{eom-background-scalars}
\ee
where $H = \dot a / a$ is the Hubble expansion rate, $V_{a} = \partial V/\partial \phi^a$, and $\gamma^{ab}$ is the inverse of $\gamma_{ab}$. In addition, $D_t$ is a covariant derivative whose action on a vector $A^a$ is given by $D_t A^a = \dot A^a + \Gamma^{a}_{bc} \dot \phi_0^b A^{c}$, where $\Gamma^{a}_{bc}$ are Christoffel symbols derived from $\gamma_{ab}$. Equation (\ref{eom-background-scalars}) must be solved together with the Friedmann equation $3 H^2 = \frac{1}{2} \dot \phi_0^2 + V(\phi_0)$, where $\dot \phi_0^2 \equiv \gamma_{ab} \dot \phi_0^a \dot \phi_0^b$ (we work in units where the reduced Planck mass is $1$). With appropriate initial conditions, this system yields a path $\phi_0^a (t)$ with tangent and normal unit vectors defined as $T^a \equiv \dot \phi^a_0 /  \dot \phi_0$ and $N^a\equiv - \frac{1}{\Omega} D_t T^a$, where $\Omega \equiv - N_a D_t T^a$ is the angular velocity with which the path bends. Successful inflation requires that $H$ stay nearly constant. This is achieved by demanding a small first slow-roll parameter $\epsilon \equiv - \dot H / H^2 \ll 1$. For simplicity, in what follows we assume that $\epsilon$ can be treated as a constant, however, we allow for an arbitrary time-dependent angular velocity $\Omega$. This corresponds to situations where the inflationary path experiences turns without affecting the expansion rate significantly. Our main conclusions will not depend on these assumptions.

We may perturb the system as $\phi^a ({\x}, t) = \phi_0^a(t) + T^a(t) \varphi  ({\x}, t) + N^a(t) \psi  (\x, t)$, where $\psi$ is the isocurvature fluctuation~\cite{Gordon:2000hv, GrootNibbelink:2001qt}. In co-moving gauge ($\varphi = 0$) we define the primordial curvature fluctuation $\zeta$ by perturbing the metric as $\de s^2 = - \mathcal N^2 \de t^2 + a^2 e^{2 \zeta} (\de {\x} + \mathbfcal{N} \de t)^2$, where $\mathcal N$ and $\mathbfcal{N}$ are the lapse and the shift. Inserting everything back into (\ref{action-starting_point}) and solving the momentum and Hamiltonian constraints, one arrives at the Lagrangian\footnote{ This Lagrangian was first derived, to second order, in Refs.~\cite{Gordon:2000hv,GrootNibbelink:2001qt}. A version with $U(\psi) \propto \psi^3$ was first considered in~\cite{Chen:2009zp} and later extended to a general potential $U(\psi)$ in~\cite{Chen:2018uul}.} $\mathcal L = \mathcal L_{\rm kin} +  \mathcal L_{\rm iso}$, with
\bea
\mathcal L_{\rm kin} =  \frac{a^3}{2} \left[ (D_t \zeta_c)^2 -   \frac{1}{a^{2}}  (\nabla \zeta_c)^2  +  \dot\psi^2  -  \frac{1}{a^{2}} (\nabla \psi)^2 \right]\! , \quad \label{Lagrangian-zeta-psi}
\eea
and $ \mathcal L_{\rm iso} = - a^3 U(\psi)$, where $U$ is a potential for $\psi$. In (\ref{Lagrangian-zeta-psi}), $\zeta_{c} \equiv \sqrt{2 \epsilon} \zeta$ is the canonically normalized version of $\zeta$, and $D_t$ is a covariant derivative  defined as
\be
D_t \zeta_c \equiv \dot \zeta_c - \lambda H \psi ,  \qquad  \lambda \equiv  2 \Omega/ H , \label{cov-D-zeta}
\ee
reminiscent of the covariant derivative in (\ref{eom-background-scalars}). The background quantity $\lambda(t)$, which is nonvanishing whenever the trajectory experiences turns, plays a prominent role in multifield inflation: it couples $\zeta$ and $\psi$ at quadratic order in a way that cannot be trivially field-redefined away~\cite{Castillo:2013sfa}. Note that in (\ref{Lagrangian-zeta-psi}) we disregarded terms that are gravitationally suppressed. For instance, we omitted terms multiplied by $\mathcal  N -1 = \dot \zeta_c / \sqrt{2 \epsilon} H$, which is much smaller than $1$ in the weak gravity regime. In what follows, we also disregard self-interactions of $\psi$ by setting $U = 0$. We will comment on the important role of $U$ toward the end of this work.


{\bf Mild mixing between $\zeta$ and $\psi$.} In the particular limit  $\lambda \ll 1$ the interaction between $\zeta$ and $\psi$ can be fully understood analytically~\cite{Achucarro:2016fby}. Here, (\ref{Lagrangian-zeta-psi}) can be split as $\mathcal L_{\rm kin} = \mathcal L_{\rm free} + \mathcal L_{\rm mix}$, where $\mathcal L_{\rm free}$ is
\be
 \mathcal L_{\rm free} = \frac{a^3}{2} \left[ \dot \zeta_c^2 - \frac{1}{a^2}  (\nabla \zeta_c)^2+  \dot\psi^2 -  \frac{1}{a^{2}} (\nabla \psi)^2 \right] , \label{L-free}
\ee
and $\mathcal L_{\rm mix}$ is the interacting part, given by
\be
 \mathcal L_{\rm mix} \equiv  - a^3 \lambda H \dot \zeta_c \psi   . \label{L-mix}
\ee
In this limit, $\zeta_c$ and $\psi$ are massless scalar fields, interacting through the mixing term proportional to $\dot \zeta_c \psi$. The Fourier space solutions of the linear equations derived from (\ref{L-free}) are
\bea
\tilde \zeta_0^c ({\K}, t) &=&  u (k, t) a_{\zeta}({\K}) +  u^* (k, t) a_{\zeta}^{\dag}(-{\K}),  \label{Fourier-solutions-1} \\
\tilde \psi_0 ({\K}, t) &=&  u (k, t) a_{\psi}({\K}) + u^* (k, t) a_{\psi}^{\dag}(-{\K}) ,  \label{Fourier-solutions-2} 
\eea
where  $a_{\zeta,\psi}({\K})$ and $a_{\zeta,\psi}^{\dag}({\K})$ are the usual creation and annihilation operators (ensuring that $\zeta$ and $\psi$ are quantum fields) satisfying $[a_a ({\K}),a_b^{\dag}({\q})] = (2 \pi)^3 \delta_{ab} \delta^{(3)}({\K} - {\q})$. Also, $u (k, t)$ is the standard de Sitter mode function respecting Bunch-Davies initial conditions:
\bea
u (t, k) =  \frac{iH}{\sqrt{2 k^3}} \left(1 + i k \tau (t) \right) e^{- i k \tau (t)} , \quad \label{mode-solutions}
\eea
where $\tau (t) = - 1/ H a(t)$ with $a(t) \propto e^{H t}$. The dimensionless power spectrum $\Delta_{\zeta} (k)$ is defined in terms of the 2-point correlation function of $\tilde \zeta$ as $\langle \tilde \zeta ({\K})  \tilde \zeta ({\q})  \rangle = (2 \pi)^3 \delta^{(3)} ({\K} - {\q}) \frac{2 \pi^2}{k^3} \Delta_{\zeta} (k)$. If $\lambda = 0$ the dynamics of $\zeta$ is of the single-field type, with a solution given by (\ref{Fourier-solutions-1}), leading to a power spectrum of the form
\be
\Delta_{\zeta,0} = \frac{H^2}{8 \pi^2 \epsilon} ,  \label{power-spectrum-zeta-0}
\ee
with small slow-roll corrections breaking its scale invariance. On the other hand, if $\lambda$ is small but nonvanishing, $\psi$ sources $\zeta_c$ via the mixing in (\ref{L-mix}) and one finds~\cite{Achucarro:2016fby}
\be
\tilde \zeta_c ({\K}) =  \tilde \zeta_0^c ({\K})  + 2 \Delta \theta (k)  \tilde \psi_0 ({\K})  , \label{zeta-psi-sourced} 
\ee
[and $\tilde \psi (\K) = \tilde \psi_0 ({\K})$], where $\Delta \theta (k) = \frac{1}{2} \int_{t_{\rm hc}}^{t_{\rm end}} \lambda H \de t$ is the total angle swept as felt by a mode that crossed the horizon at time $t_{\rm hc} = H^{-1} \ln (k/H)$. It immediately follows that $\Delta_{\zeta}$ is given by~\cite{Achucarro:2016fby}
\be
\Delta_{\zeta} = \left[ 1 + 4  \Delta \theta^2 (k) \right] \times \Delta_{\zeta,0} ,  \label{power-spectrum-zeta-lambda}
\ee
which is larger than (\ref{power-spectrum-zeta-0}) by a factor $1 + 4  \Delta \theta^2 (k)$. Equation (\ref{zeta-psi-sourced}) shows that for small $\lambda$, all superhorizon modes are equally enhanced while the turn is taking place, leading to the power spectrum (\ref{power-spectrum-zeta-lambda}) that has more amplitude on long wavelengths, independently of the form of $\lambda(t)$. This is incompatible with a large enhancement of $\Delta_{\zeta}$ on a range of wavelengths shorter than those of the CMB, forcing us to consider the regime $\lambda \gg 1$, which is the subject of the rest of this letter.


\emph{\bf Strong mixing between $\zeta$ and $\psi$.} In the case of brief turns we can study the system analytically even if $\lambda \gg 1$. Does this jeopardize the perturbativity of the system?  $\lambda$ comes exclusively from the kinetic term in (\ref{action-starting_point}) which, at quadratic order, gives rise to the covariant derivative $D_t \zeta_c$. This implies that at higher orders $\lambda$ appears in the Lagrangian through operators of order $(D_t \zeta_c)^2$ gravitationally coupled to $\zeta_c$. The condition granting that the splitting remains weakly coupled is $\mathcal L^{(3)}_{\lambda} / \mathcal L_{\lambda}^{(2)} \sim \dot \zeta_c/ \sqrt{2 \epsilon} H \ll 1$, evaluated during horizon crossing. Because at horizon crossing $(\dot \zeta_c /  \sqrt{2 \epsilon} H )^2 \sim \Delta_\zeta$, the previous requirement is equivalent to $\Delta_\zeta \ll 1$.

Hence, as long as $\Delta_\zeta \ll 1$ we may keep the mixing term into the full kinetic term of (\ref{Lagrangian-zeta-psi}). Then, the equations of motion respected by the fluctuations are
\bea
 \frac{\de}{\de t}  D_t \tilde \zeta_c  + 3 H D_t \tilde \zeta_c  + \frac{k^2}{a^2}  \tilde \zeta_c &=&  0,   \label{eq-1-full}  \\
\ddot {\tilde \psi} + 3 H \dot {\tilde \psi} + \frac{k^2}{a^2} \tilde \psi + \lambda H  D_t \tilde \zeta_c  &=& 0 . \quad \label{eq-2-full}
\eea
To proceed, let us consider the case in which $\lambda$ consists in a top-hat function of the form $\lambda (t) =  \lambda_0  \left[ \theta (t - t_1) - \theta (t - t_2) \right]$, with a small width $\delta t \equiv t_2 - t_1 \ll H^{-1}$. This profile describes a trajectory that experiences a brief turn of angular velocity $\Omega= H \lambda_0 /2$ between $t_1$ and $t_2$. Now, if $\delta t \ll H^{-1}$, during this brief period of time one can ignore the friction terms in (\ref{eq-1-full}) and (\ref{eq-2-full}), and treat the fluctuations as if they were evolving in a Minkowski spacetime~\cite{Achucarro:2010jv}. Before $t_1$, the solutions are exactly those of~(\ref{Fourier-solutions-1}), (\ref{Fourier-solutions-2}) and~\eqref{mode-solutions}. Notice that these respect Bunch-Davies initial conditions. Between $t_1$ and $t_2$ the solutions, which we denote $\tilde \Phi^a  \equiv ( \tilde \zeta_c , \tilde \psi )$, are found to be
\bea
\tilde \Phi^a ({\K}, t)  &=& \left( A^a_{\pm}  e^{ + i \omega_\pm t} +  B^a_{\pm}  e^{ - i \omega_\pm t} \right)  a_{\zeta} ({\K}) \nn \\
&& \!\!\!\!\!\!\!\!\!\!\!\!\!\!\!\!\!\!\!\!\!   + \left( C^a_{\pm}  e^{ + i \omega_\pm t} +  D^a_{\pm}  e^{ - i \omega_\pm t} \right)a_{\psi} ({\K}) + {\rm h.c.}  (-{\K}) , \qquad  \label{during-turn}
\eea
where $A^a_{\pm}$, $B^a_{\pm}$, $C^a_{\pm}$ and $D^a_{\pm}$ are amplitudes satisfying $k A^\zeta_{\pm} = \mp i \omega_{\pm} A^\psi_{\pm}$, $k B^\zeta_{\pm}= \pm i \omega_{\pm} B^\psi_{\pm}$, $k C^\zeta_{\pm} = \mp i \omega_{\pm} C^\psi_{\pm}$ and $k D^\zeta_{\pm}= \pm i \omega_{\pm} D^\psi_{\pm}$, and $\omega_{\pm}$ are the respective frequencies, given by
\be
\omega_{\pm} = \sqrt{k^2 \pm k k_0 \lambda_0}, \label{frequencies}
\ee
where $k_0 \equiv H e^{H (t_1 + t_2)/2}$ is the wave number of the modes that crossed the horizon during the turn. Finally, the solutions after $t_2$ are of the form
\bea
\tilde \Phi^a ({\K}, t) &=& \left[ E^a \, u (k, t) +  F^a \,  u^* (k, t) \right] a_{\zeta}({\K})    \nn \\ 
&& \!\!\!\!\!\!\!\!\!\!\!\!\!\!\!\!\!\!\!\!\! + \left[ G^a \, u (k, t) +  H^a \, u^* (k, t) \right] a_{\psi}({\K})   + {\rm h.c.} (-{\K}) , \qquad  \label{after-turn}
\eea
where  $u$ is given in \eqref{mode-solutions}. The amplitudes shown in Eqs.~(\ref{during-turn}) and (\ref{after-turn}) can be determined after imposing continuity of $\tilde \zeta_c ({\K}, t)$, $D_t \tilde \zeta_c ({\K}, t)$, $\tilde \psi ({\K}, t)$ and $\dot {\tilde \psi} ({\K}, t)$ at both $t_1$ and $t_2$. This is achieved by demanding the following matching conditions: 
\be
\begin{aligned}
\tilde \zeta_c ({\K}, t_1^- ) = \tilde \zeta_c ({\K}, t_1^+)  , \quad   \dot {\tilde \zeta}_c ({\K}, t_1^-) = D_t \tilde \zeta_c ({\K}, t_1^+) , \quad \\
\tilde \psi ({\K}, t_1^{-}) = \tilde \psi ({\K}, t_1^+) , \quad \dot {\tilde \psi} ({\K}, t_1^{-}) = \dot {\tilde \psi} ({\K}, t_1^+) , \quad \\
\tilde \zeta_c ({\K}, t_2^- ) = \tilde \zeta_c ({\K}, t_2^+)  , \quad D_t \tilde \zeta_c ({\K}, t_2^-) =  \dot{\tilde \zeta}_c ({\K}, t_2^+)     , \quad \\
\tilde \psi ({\K}, t_2^{-}) = \tilde \psi ({\K}, t_2^+) , \quad \dot {\tilde \psi} ({\K}, t_2^{-}) = \dot {\tilde \psi} ({\K}, t_2^+) , \quad 
\end{aligned}
\ee
where $t_i^{\pm} \equiv t_{i} \pm \varepsilon$ with $\varepsilon \to 0$. As a result, at the end of inflation, $\tilde \zeta_c ({\K})$ becomes a linear combination of quanta created and destroyed by the ladder operators $a_\zeta$ and $a_\psi$, with contributions modulated by low and high frequencies:
\bea
\tilde \zeta_c ({\K}) &=& \frac{i H}{\sqrt{2 k^3}} e^{2 i \frac{k}{k_0} \sinh \left[ \frac{\delta N}{2} \right] } \sum_{\pm} \Bigg\{  \frac{1}{2}  \bigg[   \cos \left( \frac{\omega_{\pm} \delta N}{k_0} \right) \nn \\
&&  \!\!\!\!\!\!\!\!\!\!\!\!\!\!\!
- i \frac{ k_0^2 + k^2 + \omega_{\pm}^2 }{ 2 k \omega_{\pm}}   \sin\left( \frac{\omega_{\pm} \delta N}{k_0} \right)    
\nn \\ 
&& \!\!\!\!\!\!\!\!\!\!\!\!\!\!\!
- i e^{2 i \frac{k}{k_0} \exp \left[ - \frac{\delta N}{2} \right] }  \frac{(i k_0 + k)^2 - \omega_{\pm}^2}{2 k \omega_{\pm}}   \sin \left( \frac{\omega_{\pm} \delta N}{k_0} \right) \bigg] a_{\zeta} ({\K}) \nn \\
&&  \!\!\!\!\!\!\!\!\!\!\!\!\!\!\!
\pm \frac{i}{4}  \bigg[  - \left(2 + \frac{k_0^2}{k^2} \right) \cos  \left( \frac{\omega_{\pm} \delta N}{k_0} \right) 
\nn \\
&& \!\!\!\!\!\!\!\!\!\!\!\!\!\!\! + \frac{ (k_0+ i k)k^2 -  (k_0 - i k) \omega_{\pm}^2 }{k^2 \omega_{\pm}}  \sin  \left( \frac{\omega_{\pm} \delta N}{k_0} \right)
   \qquad  \nn \\ 
 &&   \!\!\!\!\!\!\!\!\!\!\!\!\!\!\!
 + e^{2 i \frac{k}{k_0} \exp \left[ - \frac{\delta N}{2} \right] }   \bigg( \frac{k_0}{k^2} (k_0-2 i k ) \cos  \left( \frac{\omega_{\pm} \delta N}{k_0} \right)  
 \nn \\
 &&  \!\!\!\!\!\!\!\!\!\!\!\!\!\!\!
 + \frac{(k_0 -i k) (\omega_{\pm}^2 - k^2) }{k^2 \omega_{\pm} } \sin  \left( \frac{\omega_{\pm} \delta N}{k_0} \right) \bigg)  \bigg]  a_{\psi}({\K})  \Bigg\} \nn \\
 &&  \!\!\!\!\!\!\!\!\!\!\!\!\!\!\!  + \; {\rm h.c.} (-{\K}) , \label{solution-zeta-bump} 
\eea
where $\delta N \equiv H \delta t$ is the duration of the turn in $e$-folds. A similar solution can be obtained for $\tilde \psi ({\K}) $. With a WKB approximation, the same steps leading to~(\ref{solution-zeta-bump}) should allow for solutions with more general functions $\lambda (t)$. 

Equation (\ref{solution-zeta-bump}) represents our main result. It shows how $\zeta$ and $\psi$ mix after a brief turn, valid for arbitrary values of $\lambda_0$.  An outstanding feature of (\ref{solution-zeta-bump}) is that $\omega_- = \sqrt{k^2 - k k_0 \lambda_0}$ becomes imaginary for $k < \lambda_0 k_0$, signaling an instability inducing an exponential amplification of $\tilde \zeta_c ({\K})$.  On scales $k < \lambda_0 k_0$ the fluctuation displays an amplitude $\tilde \zeta_c ({\K}) \propto  e^{\sqrt{\lambda_0 k_0  k - k^2} \delta N / k_0}$ with a maximum value at $k_{\rm max} = \lambda_0 k_0 /2$. As long as $2 \delta \theta \equiv \lambda_0 \delta N > 1$ the instability generates large enhancements of the power spectrum of $\zeta$ at the end of inflation.


\emph{\bf PBHs from strong multifield mixing.} We now use the analytical insight offered by (\ref{solution-zeta-bump}) to study the origin of PBHs due to inflationary multifield effects. Direct inspection of~(\ref{solution-zeta-bump}) shows that on scales $k \gg k_0 \lambda_0 /2$,
\be
\tilde \zeta ({\K}) = i k^{-3/2} \Delta_{\zeta,0}^{1/2}  a_{\zeta} ({\K}) + {\rm h.c.} (-{\K}), 
\ee
from where one recovers the single-field power spectrum given in (\ref{power-spectrum-zeta-0}). This reveals that modes deep inside the horizon are not affected by the turn. On the other hand, on scales $k \ll k_0 \lambda_0 /2$ one has
\be
\tilde \zeta ({\K}) =  i k^{-3/2} \Delta_{\zeta,0}^{1/2} \left[ a_{\zeta} ({\K}) + 2 \delta \theta \, a_{\psi} ({\K})  \right] + {\rm h.c.}(-{\K}) ,
\ee 
implying that $\Delta_{\zeta} = \left[ 1 +  4 \delta \theta^2 \right] \times \Delta_{\zeta,0}$, confirming that on long wavelengths all scales receive the same enhancement already shown in (\ref{power-spectrum-zeta-lambda}). Finally, for $\delta \theta$ of order 1 or larger, around $k \sim k_0 \lambda_0 /2$ the curvature fluctuation is dominated by (disregarding oscillatory phases)
\bea
&& \tilde \zeta ({\K}) \sim  i k^{-3/2} \Delta_{\zeta,0}^{1/2} e^{\sqrt{\lambda_0 k_0 k - k^2} \delta N / k_0} \nn \\
&& \,\,\,\, \times \frac{1}{4}  \left[ (1 - i) a_{\zeta} (\K)  - (1 + i) a_{\psi} (\K) \right]+ {\rm h.c.} (-\K) , \label{instability} \qquad
\eea
implying that the power spectrum has a bump centered at $k = k_0 \lambda_0 /2$ of amplitude $\Delta_{\zeta,0} e^{2 \delta \theta } /4$, and width $\Delta N_k \simeq \ln[4 \delta \theta^2 / \ln^2(16\delta \theta^2)]$, where $N_k = \ln (k / H)$ is the wavenumber in $e$-fold units. In summary we have:
\be
\frac{\Delta_{\zeta} }{ \Delta_{\zeta,0}} \sim 
\left\{\begin{array}{cc}
1+ 4 \delta \theta^2 \;& \,\, {\rm if} \,\,\,  k \ll k_0 \lambda_0 /2 \\ 
\frac{1}{4} e^{2 \delta \theta }   \;& \,\, {\rm if} \,\,\, k \sim k_0 \lambda_0 /2 \\ 
1 \;&  \,\, {\rm if} \,\,\,  k \gg k_0 \lambda_0 /2 
\end{array}\right. .  \label{amplitudes-power}
\ee
This behavior is quite distinctive: for large $\delta \theta$, the ratio between the long- and short-wavelength power spectrum determines the height of the bump. In particular, the enhancement of $\Delta_\zeta$ with respect to its long-wavelength value (constrained by CMB observations) is predicted to be $\sim  e^{2\delta \theta} / (4 \delta \theta)^2$. This indicates that to have an enhancement as large as $10^{7}$, it is enough to have $\delta \theta \sim 4 \pi$.

\begin{figure}[t!]
\includegraphics[scale=0.121]{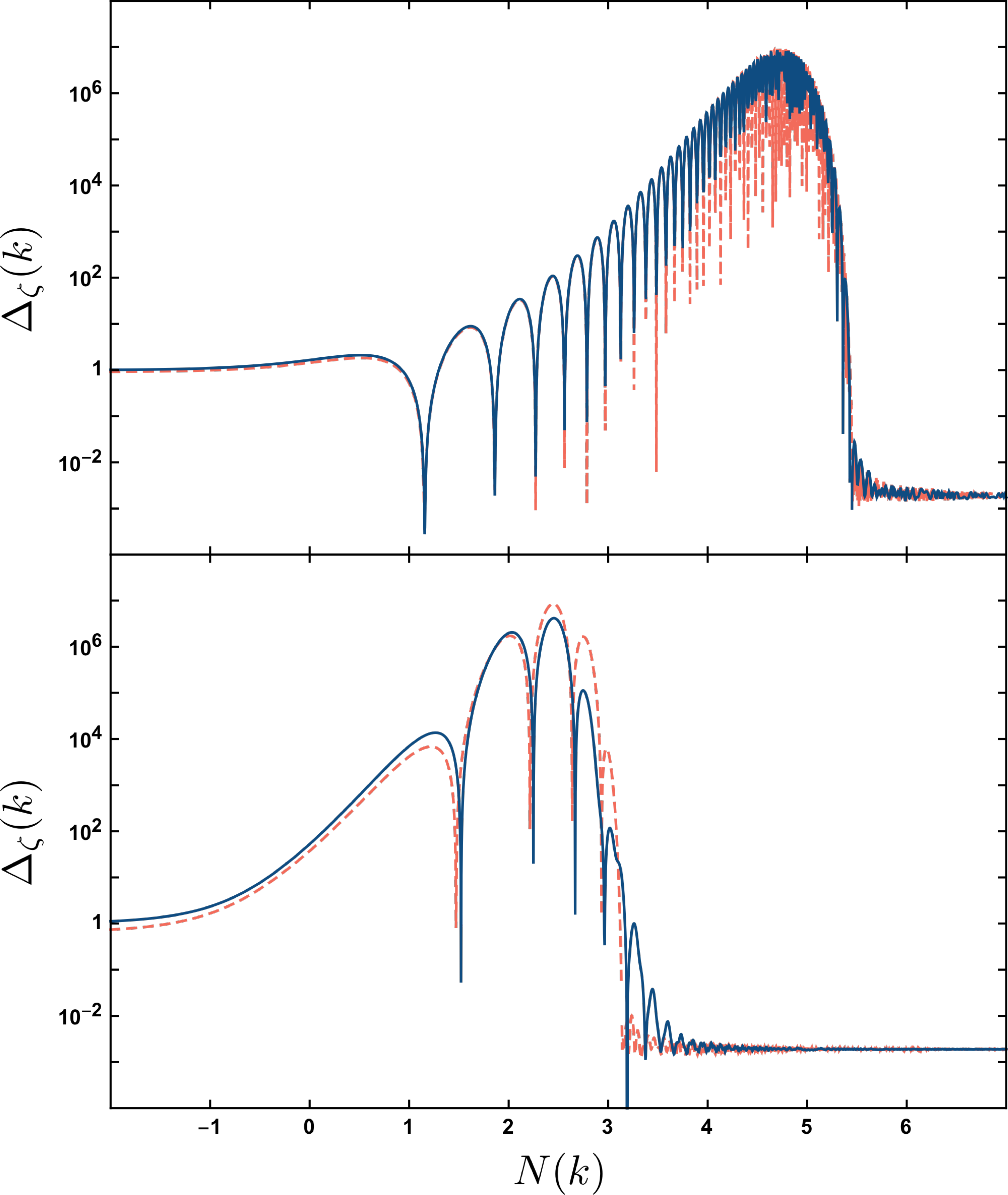}
\caption{$\Delta_\zeta (k)$ vs $N(k)=\ln (k/k_0)$ for $\delta N = 0.1$ and $\lambda_0=230$ (upper panel), and $\delta N = 1$ and $\lambda_0=23$ (lower panel). Solid/dashed lines correspond to numerical/analytical solutions. The power spectra are normalized with respect to their values at CMB scales.}
\label{fig}
\end{figure}

We can check $\Delta_\zeta (k)$ resulting from (\ref{solution-zeta-bump}) against numerical solutions of (\ref{eq-1-full}), (\ref{eq-2-full}). The results are shown in Fig.~\ref{fig} (for $\delta \theta = 11.4$). As expected, we find a good agreement for the case $\delta N = 0.1$. For $\delta N = 1$ the analytical result still constitutes a good prediction of $\Delta_{\zeta} (k)$. A salient feature of this mechanism is the rapid growth of $\Delta_\zeta(k)$ evading known limitations in achieving steep enhancements in single-field models~\cite{Byrnes:2018txb,Carrilho:2019oqg,Ozsoy:2019lyy}. Together with (\ref{solution-zeta-bump}), one can now compute the abundance of PBH as a function of their mass and the parameters $\lambda_0$ and $\delta N$, along the lines of~\cite{Germani:2018jgr,Germani:2019zez}. The power spectrum $\Delta_\zeta (k)$ displays a characteristic band structure resulting from the oscillatory phases\footnote{A similar oscillatory behavior has been observed for single-field models in Ref.~\cite{Ballesteros:2018wlw}.} in (\ref{solution-zeta-bump}). As recently studied in~\cite{Fumagalli:2020adf}, this does not necessarily translate into features in the PBHs' mass spectrum (however, our analytical results should provide further insight into understanding the observability of this band structure). Finally, one can calculate the associated gravitational wave signal, which can serve as a template for interferometric searches.


\emph{\bf A top-down example.} We now offer a concrete action of the form (\ref{action-starting_point}), where the inflationary trajectory can experience sudden turns while $\epsilon$ is kept small and constant and $U(\psi) = 0$. This can be achieved within holographic inflation~\cite{Larsen:2002et,Bzowski:2012ih,Garriga:2014fda}, where the potential $V$ in (\ref{action-starting_point}) is determined by a ``fake" superpotential $W$ as $V = 3 W^2 - 2 \gamma^{a b} W_a W_b$, with $W_a = \partial W / \partial \phi^a$. Here, the background solutions respect Hamilton-Jacobi equations:
\be
\dot \phi^a = - 2 \gamma^{a b} W_b , \qquad H = W .  \label{inflat_path} 
\ee
It turns out~\cite{Achucarro:2018ngj} that $U$ is related to $W$ as:
\bea
\partial_\psi^2 U |_{\psi = 0} &=& 6 H W_{NN} - 4 W_{NN}^2 + 2 \dot W_{NN} , \label{w-nn}
\eea
where $W_{NN} \equiv N^aN^b\nabla_aW_b$. We can now define suitable expressions for $\gamma_{ab}$ and $W$ such that $\epsilon$ and $\lambda$ have any desired time dependence. For instance, let us take fields $(\phi^1, \phi^2) = (\phi,\chi)$ and consider the following field metric:
\be
\gamma_{ab} = \left(\begin{array}{cc} e^{2 f (\phi,\chi)} & 0 \\0 & 1\end{array}\right) . \label{metric-phi-chi}
\ee
If we take a superpotential $W$ that only depends on $\phi$, then (\ref{inflat_path}) implies $\dot \phi = - 2 e^{-2f} W_\phi$ and therefore $\dot \chi = 0$ regardless of the location in field space. Thus, assuming $\dot \phi > 0$, the tangent and normal vectors are $T^a = ( e^{-f}, 0 )$ and $N^a = (0 , 1)$, and the turning rate becomes $\Omega = - N_a D_t T^a = \dot \phi \, e^f f_\chi$, implying that $\lambda H = 2 \dot \phi \, e^f f_\chi$. Also, it follows that $W_{NN} = 0$ and, thanks to (\ref{w-nn}), $\partial_\psi^2 U |_{\psi = 0}=0$. Because this result is independent of $\chi$, and $\psi$ is a perturbation along the $\chi$ direction, then $U = 0$ exactly. Next, notice that $f$ can be expanded as $f (\phi, \chi)= \sum_n \frac{1}{n!} (\chi - \chi_0)^n  f_n(\phi)$. A field redefinition of $\phi$ allows one to set $f_0(\phi)=0$. Having done so, along the specific path $\chi= \chi_0$, one finds $\dot \phi = - 2 W_\phi$ and $\lambda = \sqrt{8 \epsilon} f_1[\phi(t)]$. As a consequence, $\epsilon = 2 W_\phi^2/W^2$ and the expansion rate depends exclusively on $W(\phi)$, remaining unaffected by the turning rate $\lambda$. One can finally define $W(\phi)$ and $f_1(\phi)$ to obtain desired expressions for $\epsilon$ and $\lambda$.


\emph{\bf Discussion and conclusions.} It is well-known that sudden turns of inflationary trajectories produce features in the power spectrum~\cite{Achucarro:2010da}. Here we focused on the generation of large enhancements of the power spectrum of $\zeta$, over limited ranges of scales, compatible with PBHs confstituting a sizable fraction of dark matter.\footnote{Shortly after our paper was released Refs.~\cite{Fumagalli:2020adf,Braglia:2020eai} appeared discussing related ideas. Our conclusions, where there is overlap, are in accordance.} Noteworthily, we obtained analytical solutions for $\zeta$ valid in the regime of $\lambda \gg 1$. This is the first example of a full analytical solution in the rapid turn regime.

Large enhancements of $\Delta_{\zeta}$ demand rapid turns [recall our discussion after (\ref{power-spectrum-zeta-lambda})]. However, there is a second requirement: a nontrivial field geometry [as in (\ref{metric-phi-chi})]. For instance, consider the growth of $\Delta_{\zeta}$ in canonical multifield inflation due to a turn of duration $\delta N$ at a constant rate $\Omega$. This implies $\lambda =  2 \delta \theta / \delta N$. Nevertheless, because in canonical models $\delta \theta < \pi$, there is a maximum value $\lambda_{\rm max} = 2 \pi / \delta N$. Thus, if $\delta N \gg 1$, one falls within the slow turn regime that, as already stated, cannot produce large enhancements. On the other hand, if $\delta N\lesssim 1$, one falls within the rapid turn regime, which is well described by (\ref{amplitudes-power}), from where one learns that enhancements of order $10^7$ require $\delta \theta \sim 4 \pi$, excluding canonical models. Of course, a canonical model can still display large enhancements of $\Delta_{\zeta}$ if the expansion rate assists in amplifying $\zeta$, as found in single-field models.

How tuned is our mechanism? One of the merits of our present results is that the main parameter involved in the enhancement of the power spectrum is $\delta \theta$, the total angle swept by a turn. In that sense, our mechanism, which is exponentially sensitive to $\delta \theta$, gives a large enhancement without a large hierarchy. However, our proposal, as any other proposal based on inflation, does not necessarily provide insights as per the range of scales at which the enhancement takes place (which depends on the time at which the turn happens). Addressing this requires counting with further top-down examples such as the one offered here based on holographic inflation.

Finally, to keep the discussion simple, we ignored the potential $U$. A nonvanishing $U$ would introduce a mass for $\psi$, modifying both the dispersion relations shown in (\ref{frequencies}) and the amplitudes in (\ref{during-turn}), thus altering the details of (\ref{amplitudes-power}). Another effect is the generation of non-Gaussianity, along the lines of Refs.~\cite{Chen:2018uul,Chen:2018brw,Palma:2019lpt}. A large $\lambda$ will induce non-Gaussian deformations to the statistics of $\zeta$ that could change the details of PBH formation~\cite{Young:2013oia,Atal:2018neu,Panagopoulos:2019ail,Panagopoulos:2020sxp, Yoo:2019pma, Franciolini:2018vbk,Cai:2018dig,Passaglia:2018ixg, Atal:2019erb,Ezquiaga:2019ftu}. 

To summarize, multifield inflation can enhance the primordial power spectrum with distinctive signatures, seeding primordial black holes sensitive to the details of the ultraviolet theory wherein inflation is realized, opening a new window into early Universe physics at scales far smaller than the CMB.


\begin{acknowledgments}

\emph{Acknowledgments}: We wish to thank Vicente Atal, Kazunori Kohri, David Mulryne, Subodh Patil, Shi Pi, Walter Riquelme, Misao Sasaki, Masahide Yamaguchi and Ying-li Zhang for useful discussions and comments. GAP and CZ acknowledge support from the Fondecyt Regular Project No.~1171811 (CONICYT). SS is supported by the CUniverse research promotion project (CUAASC) at Chulalongkorn University.

\end{acknowledgments}


\end{document}